\documentclass[reprint,superscriptaddress,aps,prb]{revtex4-2}
\usepackage{dcolumn}% Align table columns on decimal point
\usepackage{bm}%
\usepackage{graphicx,graphics,color,epsfig}
\usepackage{multirow,booktabs}
\usepackage{float}
\usepackage[normalem]{ulem}  % ÓÃÓÚÉ¾³ýÏß
\usepackage{xcolor}
\usepackage{amsmath}
\newcommand\ii{\mathrm{i}}

\newcommand\nn{\nonumber}
\usepackage{hyperref}
\usepackage{amssymb}

\begin{document}

\title{Distribution of fidelity zeros in two-band topological models}

\author{Siyan Lin}
\affiliation{Beijing National Laboratory for Condensed Matter Physics, Institute
of Physics, Chinese Academy of Sciences, Beijing 100190, China}
\affiliation{School of Physical Sciences, University of Chinese Academy of Sciences,
Beijing 100049, China}

\author{Zhen-Yu Zheng}
\affiliation{Beijing National Laboratory for Condensed Matter Physics, Institute
of Physics, Chinese Academy of Sciences, Beijing 100190, China}

\author{Shu Chen}
\thanks{Corresponding author: schen@iphy.ac.cn}
\affiliation{Beijing National Laboratory for Condensed Matter Physics, Institute
of Physics, Chinese Academy of Sciences, Beijing 100190, China}
\affiliation{School of Physical Sciences, University of Chinese Academy of Sciences,
Beijing 100049, China}

\date{\today}

\begin{abstract}
    We investigate the distribution of fidelity zeros in two-band topological models by extending the phase transition driving parameter into the complex plane. Within the biorthogonal formulation, we unveil that fidelity zeros are related to momentum modes for which the real part of the energy gap vanishes. Guided by this relation, we analyze the Kitaev chain, the Haldane model, and the Qi-Wu-Zhang (QWZ) model. In finite-size systems the zeros form discrete lines parallel to the imaginary axis, while in the thermodynamic limit they accumulate into extended regions in the complex parameter plane. For the Kitaev and Haldane models, the accessible interval of the real part of the complexified parameter is bounded by the critical points of the corresponding topological transitions. For the QWZ model, the transitions at $u = \pm2$ are identified in the same way, whereas the critical point at $u = 0$ is signaled by fidelity zeros crossing the real axis. These results extend the fidelity-zero framework to topological quantum phase transitions and clarify how critical information is encoded in complexified parameter space.
\end{abstract}

\maketitle

\section{Introduction}

The study of quantum phase transitions \cite{Sachdev_2011} has always been a central topic in modern condensed matter physics. These transitions, driven by quantum fluctuations at zero temperature, signify qualitative changes in the ground-state properties of a quantum system. Therefore, the ground-state fidelity, which measures the overlap between ground states at different parameters, naturally encodes the information of quantum criticality. Fidelity and fidelity susceptibility have emerged as powerful, widely employed diagnostics for detecting quantum critical points, achieving remarkable success across a broad range of models \cite{PhysRevE.74.031123,doi:10.1142/S0217979210056335,PhysRevE.76.022101,PhysRevA.77.032111,PhysRevB.104.075142,PhysRevA.97.013845,PhysRevB.100.064427,PhysRevA.99.042117}.

Recently, inspired by Loschmidt echo zeros in dynamical quantum phase transitions \cite{PhysRevLett.110.135704,PhysRevB.89.161105,PhysRevB.89.125120,PhysRevA.97.033624,PhysRevB.102.094302,PhysRevResearch.5.033178,PhysRevB.104.094311,PhysRevB.107.134302,PhysRevA.109.043319,PhysRevA.111.042208,srx7-cpl4,qc3w-s38g}, the concept of fidelity zeros has been proposed to probe quantum phase transitions \cite{PhysRevE.109.064130}.  To overcome the Anderson orthogonality catastrophe, Ref. \cite{PhysRevE.109.064130} employed twist boundary conditions (tunable via a magnetic flux $\phi$), demonstrating that an exact fidelity zero can be accessed for states in different phases by tuning $\phi$, while no zero exists for states in the same phase. This scheme, however, operates entirely within the real parameter space of the Hamiltonian.

A more fundamental extension, in the spirit of the Lee-Yang theory \cite{PhysRev.87.404,PhysRev.87.410,PhysRevLett.93.130604,PhysRevResearch.3.033206,PhysRevB.106.054402,PhysRevResearch.5.033116,PhysRevE.110.014117,PhysRevE.109.024118,x38c-w4z2,PhysRevB.111.075134}, was subsequently developed by extending the Hamiltonian parameter into the complex plane \cite{3x34-f53v,gu2025nonhermitiansymmetrybreakingleeyang}. In this framework, fidelity zeros emerge naturally in the complex-extended parameter space, and their location obeys a Lee-Yang-type theorem, offering a unified diagnostic for quantum criticality. This formulation has been successfully applied to symmetry-breaking phase transitions such as the quantum Ising model and the clock models. In such models, fidelity zeros emerge exclusively within one phase, disappear abruptly at the critical point. Refs. \cite{3x34-f53v,gu2025nonhermitiansymmetrybreakingleeyang} attribute such behaviors to non-Hermitian parity-symmetry breaking.

However, whether this fidelity-zero framework can be applied to topological systems remains an open question. Topological quantum phase transitions involve changes in global topological invariants without local order parameters or spontaneous symmetry breaking. The mechanism of non-Hermitian symmetry breaking, central to prior demonstrations, is not directly applicable to topological systems. It is interesting to scrutinize topological quantum phase transitions in the fidelity-zero framework and clarify how critical information is encoded in complexified parameter space.

In this work, we study fidelity zeros in two-band topological models. We demonstrate that fidelity zeros emerge in the complex parameter plane precisely when the real part of the energy gap closes for specific momentum modes. Using the Kitaev chain, the Haldane model, and the Qi-Wu-Zhang (QWZ) model as paradigmatic examples, we investigate the distribution of fidelity zeros and find these zeros display similar patterns: these zeros are distributed along lines parallel to the imaginary axis in finite-size systems, with their real parts confined to a finite interval. Importantly, as the system approaches the thermodynamic limit, the boundaries of this interval converge and ultimately correspond to the critical points of the topological quantum phase transition. We establish that the presence or absence of fidelity zeros reliably signals the topological phase boundary, indicating that the fidelity-zero approach offers a powerful tool for probing topological quantum phase transitions.

\section{Fidelity zeros and their connection to the energy spectrum}

Fidelity zeros are analogous to Lee-Yang zeros, which are distributed in the complex plane. To explore such zeros, it is necessary to extend at least one parameter of the Hamiltonian into the complex domain, thereby rendering the Hamiltonian non-Hermitian. The non-Hermiticity leads to two important consequences. First, since the eigenenergies are generally complex, the concept of a ``ground state'' is defined with respect to the real part of the energy; the state with the smaller real part is identified as the lower-energy state. Second, the calculation of fidelity should be formulated using the biorthogonal basis \cite{Brody_2014} constructed from the left and right eigenstates.

We consider a generic two-band system with the momentum-space Hamiltonian expressed as
\begin{equation}
    H_k(\gamma)=d_k^0(\gamma)\mathbb{I}_2+\sum_{\beta=1}^3{d_k^\beta(\gamma)\sigma_\beta},
    \label{eq:2-band Hamiltonian}
\end{equation}
where $\sigma_\beta~(\beta=1,2,3)$ is the Pauli matrix, $\mathbb{I}_2$ is the $2\times2$ identity matrix, and $\gamma$ is a phase transition driving parameter.
The energy spectrum of this two-band model is given by
\begin{align}
    E_{k,\pm}(\gamma)=d_k^0(\gamma)\pm\sqrt{\sum_{\beta=1}^3{d_k^\beta(\gamma)^2}},
\end{align}
and the corresponding right eigenstates can be denoted as
\begin{align}
    |\Psi^R_{k,\pm}(\gamma)\rangle=\frac{1}{\sqrt{\left|d_k^3+E_{k,\pm}\right|^2+\left|d_k^1+\ii d_k^2\right|^2}}
    \begin{pmatrix}
        d_k^3+E_{k,\pm} \\ d_k^1+\ii d_k^2
    \end{pmatrix}.
\end{align}
Due to the complex nature of the spectrum, it is not \textit{a priori} evident whether $|\Psi^R_{k,+}(\gamma)\rangle$ or $|\Psi^R_{k,-}(\gamma)\rangle$ corresponds to the lower-energy state. We therefore denote the right eigenstate with the smaller (larger) real part of energy as $|\Psi^R_{k,g(e)}(\gamma)\rangle$. Similarly, we introduce the left eigenstates satisfying
\begin{align}
H_k^\dagger(\gamma)|\Psi^L_{k,\pm}(\gamma)\rangle=E_{k,\pm}^*(\gamma)|\Psi^L_{k,\pm}(\gamma)\rangle,
\end{align}
and denote the lower- (higher-) energy left eigenstate as $|\Psi^L_{k,g(e)}(\gamma)\rangle$. The left and right eigenstates are normalized as
\begin{align}
    \langle\Psi^{R}_{k,\pm}(\gamma)|\Psi^{R}_{k,\pm}(\gamma)\rangle=\langle\Psi^{L}_{k,\pm}(\gamma)|\Psi^{R}_{k,\pm}(\gamma)\rangle=1.
\end{align}

Given the non-Hermitian nature of the Hamiltonian, the fidelity is evaluated within the biorthogonal framework \cite{PhysRevA.98.052116}. Consequently, the fidelity factorizes over momentum modes as \cite{Sun2021,PhysRevResearch.3.013015}
\begin{align}
    \mathcal{F}(\tilde{\gamma},\gamma)=\prod_k\sqrt{\left|\langle\Psi^L_{k,g}(\tilde{\gamma})|\Psi^R_{k,g}(\gamma)\rangle\langle\Psi^L_{k,g}(\gamma)|\Psi^R_{k,g}(\tilde{\gamma})\rangle\right|}.
    \label{eq:fidelity}
\end{align}

In the thermodynamic limit, the many-body fidelity always approaches zero, which can be attributed to the Anderson orthogonality catastrophe \cite{PhysRevLett.18.1049,PhysRev.164.352}. To reliably identify genuine zeros in large but finite systems, we therefore introduce the minimum single-mode fidelity
\begin{align}
    &\mathcal{F}_{\text{min}}(\tilde{\gamma},\gamma)\nn\\
    =&\min_{k}\left[\sqrt{\left|\langle\Psi^L_{k,g}(\tilde{\gamma})|\Psi^R_{k,g}(\gamma)\rangle\langle\Psi^L_{k,g}(\gamma)|\Psi^R_{k,g}(\tilde{\gamma})\rangle\right|}\right],
    \label{eq:Fmin}
 \end{align}
where $k$ takes discrete values for a finite-size system. This quantity isolates the momentum mode that contributes the smallest overlap. Since the total many-body fidelity is a product over all momentum modes [Eq.~\eqref{eq:fidelity}], the vanishing of a single factor suffices to force the total fidelity to zero. Thus, a zero in $\mathcal{F}_{\text{min}}$ serves as a precise indicator for genuine zeros of the total many-body fidelity.

We now elucidate the fundamental connection between the emergence of fidelity zeros and the spectral properties of the non-Hermitian Hamiltonian. A fidelity zero between two infinitesimally close parameters, $\mathcal{F}(\gamma+\delta\gamma/2,\gamma-\delta\gamma/2)\to0$, signals a non-analyticity in the ground state at the point $\gamma$. This non-analyticity is directly triggered when, for some momentum $k$, the real part of the energy gap closes, i.e., $\operatorname{Re}\left[E_{k,+}(\gamma)-E_{k,-}(\gamma)\right]=0$. At such a point, denoted as $\gamma_0$, the two eigenstates $|\Psi_{k,\pm}^{R(L)}(\gamma_0)\rangle$ become degenerate in their real part of the energy. Consequently, the identification of the ground-state changes discontinuously as the parameter crosses $\gamma_0$. The state with the lower real part switches from, for instance, $|\Psi_{k,-}^{R(L)}\rangle$ (for $\gamma=\gamma_0-\delta\gamma/2$) to $|\Psi_{k,+}^{R(L)}\rangle$ (for $\gamma=\gamma_0+\delta\gamma/2$). The fidelity factor for this momentum mode, evaluated between states on opposite sides of $\gamma_0$, therefore involves the overlap
\begin{widetext}\begin{align}
    &\langle\Psi_{k,g}^L(\gamma_0+\delta\gamma/2)|\Psi_{k,g}^R(\gamma_0-\delta\gamma/2)\rangle\langle\Psi_{k,g}^L(\gamma_0-\delta\gamma/2)|\Psi_{k,g}^R(\gamma_0+\delta\gamma/2)\rangle\nn\\
    =&\langle\Psi_{k,+}^L(\gamma_0+\delta\gamma/2)|\Psi_{k,-}^R(\gamma_0-\delta\gamma/2)\rangle\langle\Psi_{k,-}^L(\gamma_0-\delta\gamma/2)|\Psi_{k,+}^R(\gamma_0+\delta\gamma/2)\rangle.
\end{align}\end{widetext}
In the limit $\delta\gamma\to0$, this overlap approaches $\langle\Psi_{k,+}^L(\gamma_0)|\Psi_{k,-}^R(\gamma_0)\rangle\langle\Psi_{k,-}^L(\gamma_0)|\Psi_{k,+}^R(\gamma_0)\rangle$, which is zero due to the biorthogonality of the eigenstates. Therefore, we demonstrate that a genuine fidelity zero emerges precisely when the real part of the energy gap closes, i.e., $\operatorname{Re}\left[E_{k,+}(\gamma)-E_{k,-}(\gamma)\right]=0$, for a specific momentum mode $k$.

Based on this intrinsic connection between fidelity zeros and the energy spectrum, we introduce the minimum real part of the energy gap
\begin{align}
    E_{\text{min}}(\gamma)=\min_{k\in\text{1BZ}}\left|\operatorname{Re}(E_{k,+}-E_{k,-})\right|,
    \label{eq:Emin}
\end{align}
where $k$ takes values in the first Brillouin zone (1BZ) continuously. A necessary condition for the occurrence of a fidelity zero at some $\gamma=\gamma_0$ is that $E_{\text{min}}(\gamma_0)=0$, i.e., the real part of the energy gap must vanish for at least one momentum mode. However, we emphasize that the converse is not automatically guaranteed in a finite-size system. Because the momenta are discrete in a finite-size system, e.g., $k=2\pi m/L$ with $m=0,1,\cdots,L/2$ for the Kitaev chain, the actual minimum over the discrete set, $\displaystyle\min_{\text{discrete }k}{\left|\operatorname{Re}(E_{k,+}-E_{k,-})\right|}$, may remain nonzero even at parameters where $E_{\text{min}}(\gamma)=0$ holds. Consequently, while the condition $E_{\text{min}}(\gamma)=0$ gives the locations of fidelity zeros in the thermodynamic limit, it is a necessary but not sufficient condition for their appearance in finite-size systems.

\section{Distribution of fidelity zeros}

In this section, we investigate the distribution of fidelity zeros for some typical topological models via the minimum real part of the energy gap [Eq. \eqref{eq:Emin}] and the minimum single-mode fidelity [Eq. \eqref{eq:Fmin}].

\subsection{Kitaev chain}
First, we consider the one-dimensional Kitaev chain \cite{Kitaev_2001}, a paradigmatic model for topological superconductivity. Its Hamiltonian can be cast into the standard two-band form of Eq. \eqref{eq:2-band Hamiltonian} with the components
\begin{align}
    d^0_k=d^1_k=0,\quad d^2_k=-\Delta\sin{k},\quad d^3_k=-\mu-t\cos{k},
\end{align}
where $t$ is the hopping amplitude, $\mu$ is the on-site chemical potential, and $\Delta$ denotes the amplitude of p-wave superconducting pairing. For simplicity, we set $t=1$. The energy spectrum is given by
\begin{align}
    E_{k,\pm}(\mu)=\pm\sqrt{{\left(\mu+\cos{k}\right)^2+\left(\Delta\sin{k}\right)^2}}.
\end{align}
For a system of size $L$, the allowed momenta are given by $k=2\pi m/L$ with $m=0,1,\cdots,L/2$, where we assume $L$ to be even.

The topology of the Kitaev chain is characterized by the winding number $\nu$. The model undergoes topological quantum phase transitions at
\begin{align}
    \mu = \pm 1 \quad \text{(for } \Delta \neq 0\text{)}.
\end{align}
The system resides in a topologically non-trivial phase with $\nu=1$ for $\left|\mu\right|<1$, and in a topologically trivial phase with $\nu=0$ for $\left|\mu\right|>1$.

\begin{figure}
    \centering
    \includegraphics[width=\linewidth]{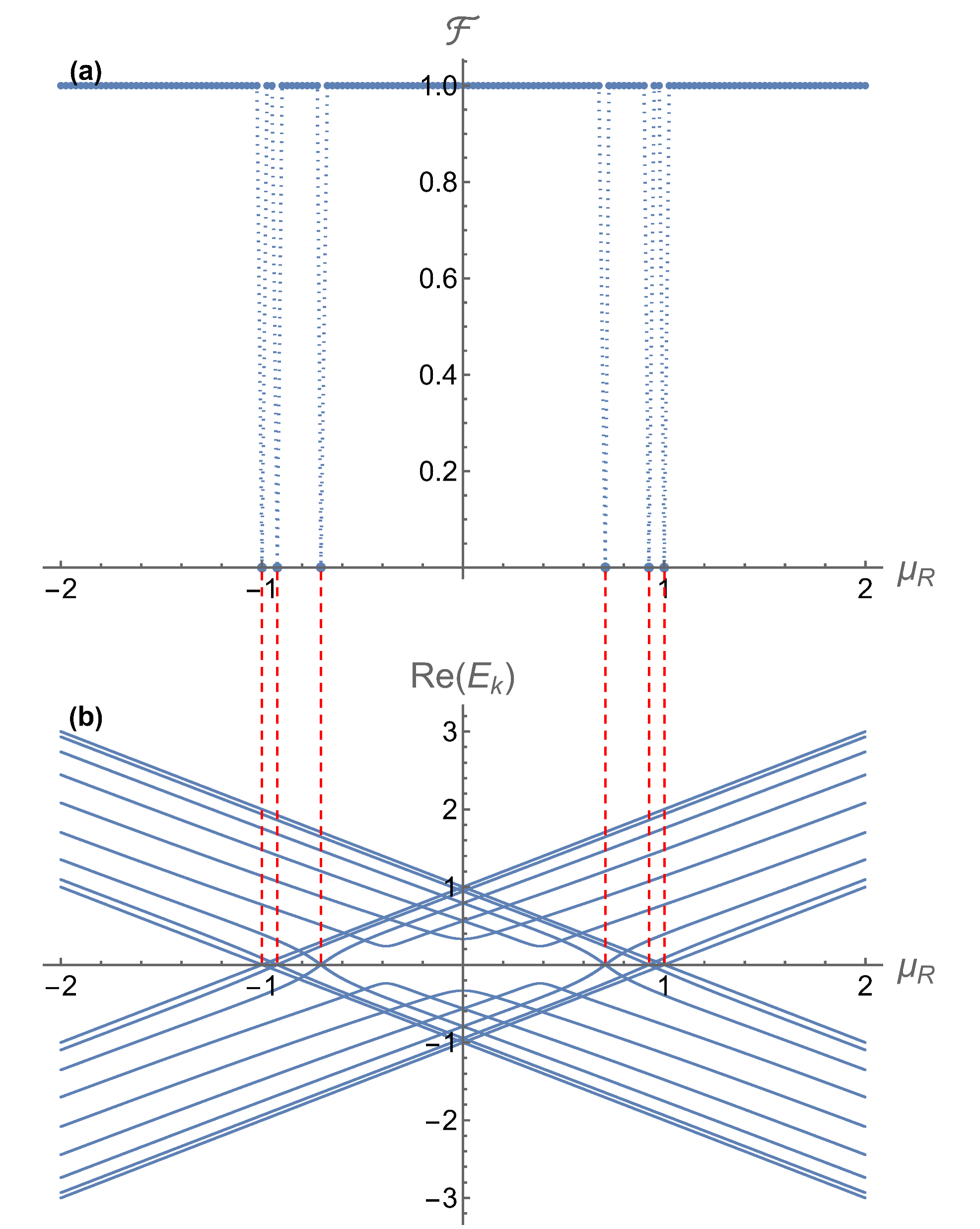}
    \caption{(a) Fidelity $\mathcal{F}(\mu+\delta\mu/2,\mu-\delta\mu/2)$ and (b) the real part of the eigenenergies $\operatorname{Re}(E_{k,\pm})$ as a function of $\mu_R$, with fixed $\Delta=0.6,~\mu_I=0.5,~\delta\mu=0.001(1+\ii)$ and $L=16$. The positions where $\operatorname{Re}(E_{k,\pm})$ vanishes coincide precisely with the fidelity zeros, as indicated by the red dashed lines.}
    \label{fig:F & ReE}
\end{figure}

\begin{figure}
    \centering
    \includegraphics[width=.8\linewidth]{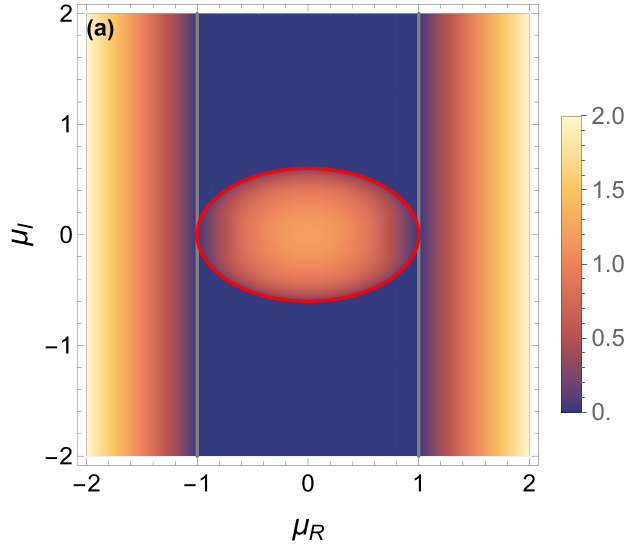}
    \includegraphics[width=.8\linewidth]{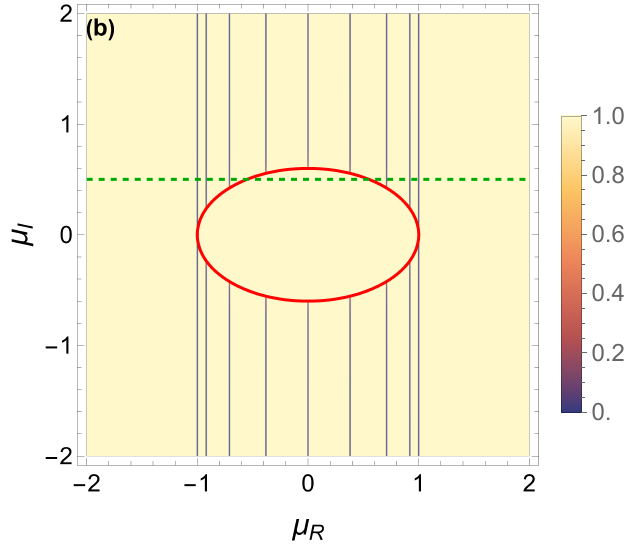}
    \caption{(a) The minimum real part of the energy gap $E_{\text{min}}(\mu)$ and (b) the minimum single-mode fidelity $\mathcal{F}_{\text{min}}(\mu+\delta\mu/2,\mu-\delta\mu/2)$ as a function of $\mu$ in the complex plane for the Kitaev chain. We fix $\Delta=0.6$ for both panels. In panel (b) we fix $\delta\mu=0.01(1+\ii)$ and $L=16$. }
    \label{fig:Kitaev}
\end{figure}

To explicitly verify the connection between fidelity zeros and the energy spectrum, we extend the chemical potential into the complex plane, setting $\mu=\mu_R+\ii\mu_I$. We then compute both the ground-state fidelity $\mathcal{F}(\mu+\delta\mu/2,\mu-\delta\mu/2)$ and the real part of the eigenenergies $\operatorname{Re}(E_k)$ along a chosen path in the complex $\mu$-plane. Here we fix $\mu_I=0.5$ and vary $\mu_R$ [see the green dashed line in Fig. \ref{fig:Kitaev}(b)]. As shown in Fig. \ref{fig:F & ReE}(a), the fidelity remains a constant value of 1 except at some specific parameter values, where sharp dips or zeros of the fidelity are clearly visible. These zeros coincide precisely with the points where the real part of the energy gap closes, as indicated by the red dashed lines in Fig. \ref{fig:F & ReE}. This one-to-one correspondence confirms the argument presented earlier: a fidelity zero emerges if and only if the real part of the energy gap closes for at least one momentum mode. Besides, fidelity zeros are confined within the interval $\left|\mu_R\right|\le1$. The boundaries of this interval give rise to the topological quantum phase transition points.

Next, we examine the distribution of fidelity zeros in the complex $\mu$-plane. In Fig. \ref{fig:Kitaev}(a) we plot the minimum real part of the energy gap $E_{\text{min}}(\mu)$ for $\Delta=0.6$. It is shown that there exists a region where $E_{\text{min}}(\mu)=0$, with boundaries indicated by the red closed curve and gray lines. Importantly, the gray lines are located at $\left|\mu_R\right|=1$, coinciding with the critical points of the underlying topological quantum phase transitions. For $\left|\mu_R\right|>1$, $E_{\min}(\mu)$ is strictly positive, indicating the absence of fidelity zeros in that region. The red curve can be determined analytically by solving
\begin{align}
    E_{k,\pm}(\mu)=\pm\sqrt{{\left(\mu+\cos{k}\right)^2+\left(\Delta\sin{k}\right)^2}}=0,
\end{align}
which yields the parametric form
\begin{align}
    \mu=-\cos{k}\pm\ii\Delta\sin{k}.
    \label{eq:parametric mu}
\end{align}
Consequently, the region where $E_{k,\pm}(\mu)=0$ is given by
\begin{align}
    R=\left\{\mu=\mu_R+\ii\mu_I:\left|\mu_R\right|\leq1,~ \mu_R^2+\left(\frac{\mu_I}{\Delta}\right)^2\ge1\right\}.
\end{align}

We now turn to the numerical detection of fidelity zeros, using the minimum single-mode fidelity $\mathcal{F}_{\text{min}}$ defined in Eq. \eqref{eq:Fmin}. In Fig. \ref{fig:Kitaev}(b) we plot $\mathcal{F}_{\text{min}}$ in the complex $\mu$-plane. The dark blue lines are identified as the fidelity zeros. These fidelity zeros lie in the region $R$ as expected, and are distributed along lines parallel to the imaginary axis. The distribution of fidelity zeros can be obtained analytically, by requiring the real part of the energy gap to vanish, i.e. $\operatorname{Re}(E_{k,+}(\mu)-E_{k,-}(\mu))=0$. This is equivalent to the condition
\begin{align}
    (\mu+\cos{k})^2+(\Delta\sin{k})^2\leq0.
\end{align}
Substituting $\mu=\mu_R+\ii\mu_I$ into this inequality and expanding the left-hand side, we obtain
\begin{align}
    \mu_R^2&-\mu_I^2+\cos^2{k}+2\mu_R\cos{k}+(\Delta\sin{k})^2\nn\\
    &+2\ii(\mu_R+\cos{k})\mu_I\leq0.
\end{align}
The real part and imaginary part give two separate constraints
\begin{align}
    \mu_R^2-\mu_I^2+\cos^2{k}+2\mu_R\cos{k}+(\Delta\sin{k})^2&\leq0,\nn\\
    2(\mu_R+\cos{k})\mu_I&=0.
\end{align}
For a finite-size system with discrete momenta $k=2\pi m/L~(m=0,1,\cdots,L/2)$, these constraints yields the explicit locations of fidelity zeros
\begin{subequations}\begin{align}
    &\mu_R=-\cos\left(\frac{2\pi m}{L}\right)\quad \left(m=0,1,\cdots,\frac{L}{2}\right),\\
    &\mu_R^2+\left(\frac{\mu_I}{\Delta}\right)^2\geq1,
\end{align}\end{subequations}

Crucially, the occurrence of these zeros depends sharply on the real part $\mu_R$ of the complexified parameter. For $\left|\mu_R\right|>1$, no zeros appear, whereas for $\left|\mu_R\right|\leq1$ zeros are present. This abrupt change in the zero pattern precisely at $\left|\mu_R\right|=1$ provides a strong indication of the topological quantum phase transition in the original Hermitian Kitaev chain, where the critical points are located at $\mu=\pm1$. Thus, monitoring the existence of fidelity zeros as a function of $\mu_R$ offers a diagnostic of the underlying quantum phase transition.

Moreover, we note that the Su-Schrieffer-Heeger (SSH) model \cite{PhysRevLett.42.1698,PhysRevB.22.2099} can be viewed as a special case of the Kitaev chain with $\Delta=t$ after a suitable rotation of the vector $\boldsymbol{d}_k$. The Hamiltonian of the SSH model takes the form of Eq. \eqref{eq:2-band Hamiltonian} with the components
\begin{align}
    d^0_k=d^3_k=0,\quad d^1_k=t_1+t_2\cos{k},\quad d^2_k=t_2\sin{k},
\end{align}
where $t_1$ and $t_2$ are intra‑cell and inter‑cell hopping amplitudes, respectively. By performing a $\pi/2$ rotation around the $d^2_k$-axis and identifying $\mu=-t_1,~\Delta=t=-t_2$, one recovers the Kitaev chain. Consequently, the analysis of fidelity zeros for the SSH model follows directly from the Kitaev chain results. For simplicity, we fix $t_2=-1$ and extend $t_1$ into the complex plain, $t_1=t_{1R}+\ii t_{1I}$. Then fidelity zeros are located at
\begin{subequations}\begin{align}
    &t_{1R}=\cos\left(\frac{2\pi m}{L}\right),\quad \left(m=-\frac{L}{2}+1,-\frac{L}{2}+2,\cdots,\frac{L}{2}\right)\label{eq:t_1R}\\
    &t_{1R}^2+t_{1I}^2\ge1. 
\end{align}\end{subequations}
From Eq. \eqref{eq:t_1R} we can directly deduce $\left|{t_{1R}}\right|\leq1$. The boundaries exactly give rise to the critical points of the SSH model. 

\subsection{Haldane model}

Next, we examine fidelity zeros in the Haldane model on a honeycomb lattice \cite{PhysRevLett.61.2015}, a paradigmatic example of a Chern insulator that exhibits the quantum anomalous Hall effect. Its Bloch Hamiltonian can be written in the standard two-band form of Eq. \eqref{eq:2-band Hamiltonian} with the components
\begin{widetext}\begin{align}
    d^0_{\boldsymbol{k}}&=2t_2\cos{\theta}\left[\cos\left(\sqrt{3}k_x\right)+2\cos\left(\frac{\sqrt{3}}{2}k_x\right)\cos\left(\frac{3}{2}k_y\right)\right],\qquad
    d^1_{\boldsymbol{k}}=t_1\left[\cos{k_y}+2\cos\left(\frac{\sqrt{3}}{2}k_x\right)\cos\left(\frac{k_y}{2}\right)\right],\nn\\
    d^2_{\boldsymbol{k}}&=t_1\left[\sin{k_y}-2\cos\left(\frac{\sqrt{3}}{2}k_x\right)\sin\left(\frac{k_y}{2}\right)\right],\qquad
    d^3_{\boldsymbol{k}}=M-2t_2\sin{\theta}\left[\sin\left(\sqrt{3}k_x\right)-2\sin\left(\frac{\sqrt{3}}{2}k_x\right)\cos\left(\frac{3}{2}k_y\right)\right],
\end{align}\end{widetext}
where $t_1$ and $t_2$ are nearest-neighbor and next-nearest-neighbor hopping amplitudes, $\theta$ is an additional effective phase of hopping between next-nearest-neighbor sites, and $M$ is the staggered sublattice potential. For a finite lattice of size $N=L\times L$, the momenta are quantized as $k_x=\frac{2\pi m_x}{\sqrt{3}L},~k_y=\frac{4\pi m_y}{3L}$ with $m_{x(y)}=1,2,\cdots,L$. The model is characterized by the Chern number $\mathcal{C}$. It undergoes topological quantum phase transitions at
\begin{align}
    M=\pm3\sqrt{3}t_2\sin{\theta},
\end{align}
which separate a topologically non-trivial phase with $\mathcal{C}=\pm1$ from a topologically trivial phase with $\mathcal{C}=0$. For simplicity we set $t_1=1,~t_2=1/2$ and $\theta=\pi/6$ in the following discussion. The critical points then lie at $M=\pm\frac{3\sqrt{3}}{4}\approx\pm1.30$.

\begin{figure}
    \centering
    \includegraphics[width=.8\linewidth]{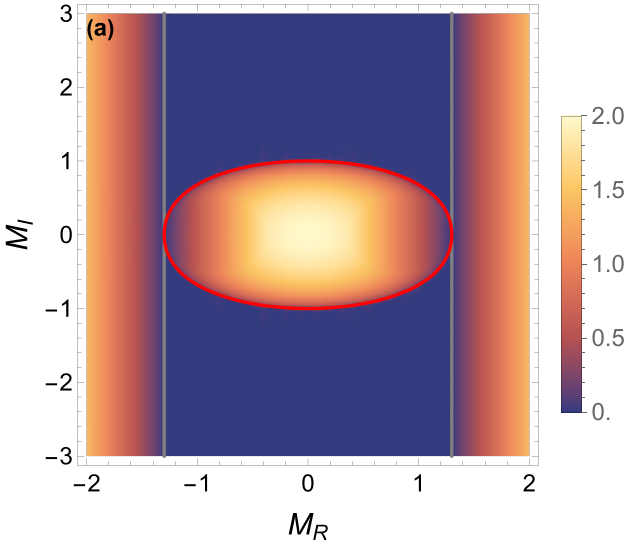}
    \includegraphics[width=.8\linewidth]{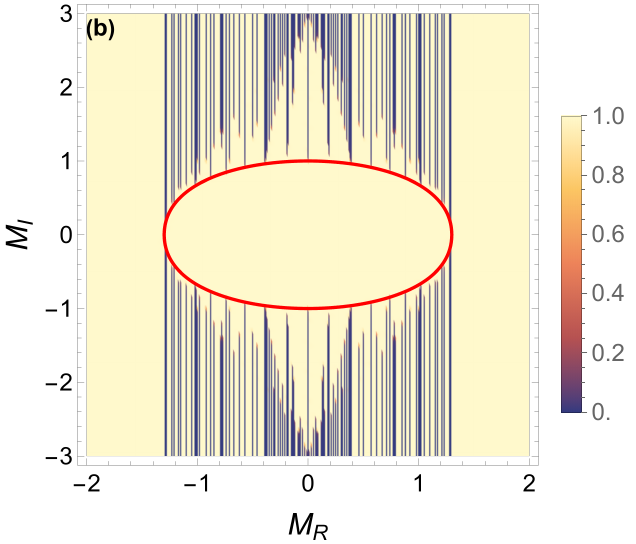}
    \caption{(a) The minimum real part of the energy gap $E_{\text{min}}(M)$ and (b) the minimum single-mode fidelity $\mathcal{F}_{\text{min}}(M+\delta M/2,M-\delta M/2)$ as a function of $M$ in the complex plane for the Haldane model. In panel (b) we fix $\delta M=0.01(1+\ii)$ and $L=16$. }
    \label{fig:Haldane}
\end{figure}

In order to explore fidelity zeros in the complex plane, we extend $M$ into the complex domain $M=M_R+\ii M_I$. We first compute the minimum real part of the energy gap $E_{\text{min}}$ as a function of $M$ in the complex plane, as shown in Fig. \ref{fig:Haldane}(a). A clear structural pattern similar to the case of Kitaev chain emerges. For $\left|M_R\right|\le1.30$ there is a region where $E_{\text{min}}(M)=0$, with boundaries indicated by the red closed curve and gray lines. Within this region, the real part of the energy gap closes for at least one momentum mode, satisfying the necessary condition for the appearance of fidelity zeros. In contrast, for $\left|M_R\right|>1.30$, $E_{\text{min}}(M)$ remains strictly positive, indicating that no spectral degeneracy in the real part occurs and hence fidelity zeros are absent. This sharp change in the behavior of $E_{\text{min}}(M)$ across $\left|M_R\right|\approx1.30$ again provides a direct link between the complex?plane zeros and the underlying quantum phase transition of the original Hermitian system.

For a finite-size system, we also analyze the distribution of fidelity zeros via the minimum single?mode fidelity defined in Eq. \eqref{eq:Fmin}. Figure \ref{fig:Haldane}(b) shows $\mathcal{F}_{\text{min}}(M+\delta M/2,M-\delta M/2)$ as a function of $M$, with fixed $\delta M=0.01(1+\ii)$. The dark blue lines are identified as fidelity zeros. These zeros are distributed along lines parallel to the imaginary axis. The exact positions of fidelity zeros can also be obtained analytically by requiring the energy gap to be pure imaginary, i.e., $\operatorname{Re}(E_{\boldsymbol{k},+}-E_{\boldsymbol{k},-})=0$. Solving this condition yields explicit constraints on the complex parameter $M=M_R+\ii M_I$
\begin{subequations}\begin{align}
    M_R&=-4t_2\sin\left(\frac{\sqrt{3}}{2}k_x\right)\sin{\theta}\left[\cos\left(\frac{3}{2}k_y\right)-\cos\left(\frac{\sqrt{3}}{2}k_x\right)\right],\label{eq:M_R}\\
    M_I^2&\ge t_1^2\left[3+2\cos\left(\sqrt{3}k_x\right)+4\cos\left(\frac{\sqrt{3}}{2}k_x\right)\cos\left(\frac{3}{2}k_y\right)\right].
\end{align}\label{eq:Haldane FZ}\end{subequations}
These constraints provide the locus of fidelity zeros for a given momentum $\boldsymbol{k}$. We verify these analytic expressions in Appendix \ref{appendix}. Moreover, from the explicit expression for $M_R$ in Eq. \eqref{eq:M_R}, we can directly deduce its permissible range $\left|M_R\right|\le3\sqrt{3}t_2\sin{\theta}$. This derived bound coincides precisely with the location of the topological quantum phase transition points in the original Hermitian Haldane model. This again indicates that for the Haldane model, fidelity zeros can appear only when the real part of the complexified parameter lies inside one phase (the topologically non-trivial phase) of the original Hermitian model.

\subsection{QWZ model}
Finally, we consider the QWZ model \cite{PhysRevB.74.085308}, which is a two-dimensional two-band model described by the Hamiltonian Eq. \eqref{eq:2-band Hamiltonian} with components
\begin{align}
    d_{\boldsymbol{k}}^0&=0,\qquad d_{\boldsymbol{k}}^1=\sin{k_x},\qquad d_{\boldsymbol{k}}^2=\sin{k_y},\nn\\
    d_{\boldsymbol{k}}^3&=u+\cos{k_x}+\cos{k_y},
\end{align}
where $u$ is the staggered onsite potential. Depending on the value of $u$, this model has three different phases characterized by the Chern number $\mathcal{C}$: $\left|u\right|>2$ with $\mathcal{C}=0$, $-2<u<0$ with $\mathcal{C}=-1$, and $0<u<2$ with $\mathcal{C}=1$. The energy spectrum is given by
\begin{align}
    E_{\boldsymbol{k},\pm}=\pm\sqrt{\sin^2{k_x}+\sin^2{k_y}+\left(u+\cos{k_x}+\cos{k_y}\right)^2}.
\end{align}
For a system of size $N=L\times L$, the allowed momenta are given by $k_{x(y)}=\frac{2\pi m_{x(y)}}{L}$ with $m_{x(y)}=1,2,\cdots,L$.

\begin{figure}
    \centering
    \includegraphics[width=.8\linewidth]{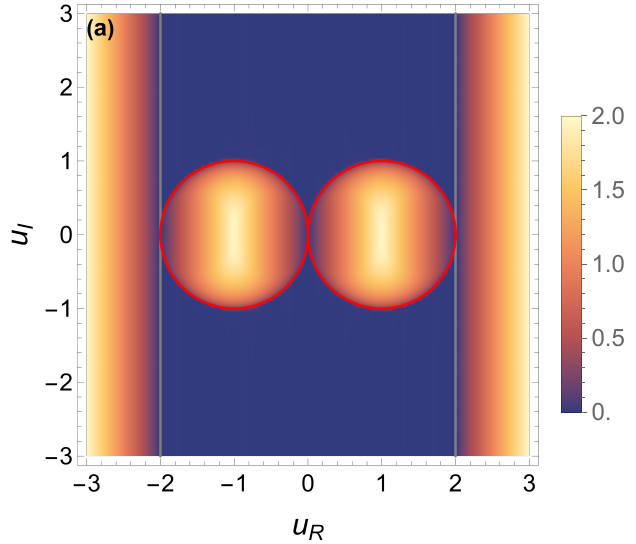}
    \includegraphics[width=.8
    \linewidth]{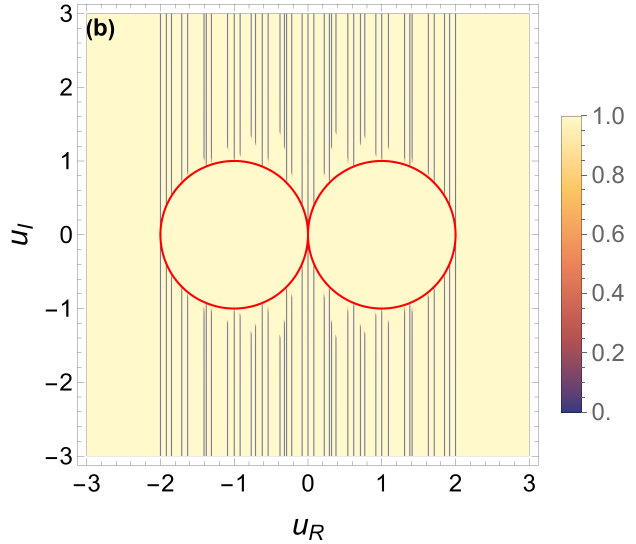}
    \caption{(a) The minimum real part of the energy gap $E_{\text{min}}(\mu)$ and (b) the minimum single-mode fidelity $\mathcal{F}_{\text{min}}(u+\delta u/2,u-\delta u/2)$ as a function of $u$ in the complex plane for the QWZ model. In panel (b) we fix $\delta u=0.01(1+\ii)$ and $L=16$. }
    \label{fig:QWZ}
\end{figure}

We first compute the minimum real part of the energy gap $E_{\text{min}}(u)$ of the QWZ model in the complex $u$-plane, as shown in Fig. \ref{fig:QWZ}(a). For $\left|u_R\right|\le2$, a region exhibiting $E_{\text{min}}(u)=0$ emerges, with red closed curves and gray lines indicating the boundaries. For $\left|u_R\right|>2$, $E_{\text{min}}(u)$ remains strictly positive, indicating the absence of fidelity zeros. Again, monitoring the existence of fidelity zeros as a function of $u_R$ offers a diagnostic of the topological quantum phase transition at $u=\pm2$. The critical point at $u=0$ cannot be directly extracted from the existence or non-existence of fidelity zeros. However, fidelity zeros will cross the real axis at $u=0$ in the thermodynamic limit, since the two red closed curves are tangent to each other at $u=0$. Following the philosophy of the Lee-Yang theory, the zero approaching the real axis represents the critical point.

Next, we investigate the distribution of fidelity zeros for the QWZ model via the minimum single-mode fidelity. In Fig. \ref{fig:QWZ}(b) we plot $\mathcal{F}_{\text{min}}(u+\delta u/2,u-\delta u/2)$ as a function of $u$, with fixed $\delta u=0.01(1+\ii)$. These zeros again lie along lines parallel to the imaginary axis. We derive the exact positions of fidelity zeros, by requiring $\operatorname{Re}\left(E_{\boldsymbol{\boldsymbol{k},+}}-E_{\boldsymbol{k},-}\right)=0$. Solving this condition yields explicit constraints on the complex parameter $u=u_R+\ii u_I$
\begin{subequations}\begin{align}
    u_R&=-\cos{k_x}-\cos{k_y},\label{eq:u_R}\\
    u_I^2&\ge2-\cos^2{k_x}-\cos^2{k_y}.
\end{align}\label{eq:QWZ FZ}\end{subequations}
Verification of these constraints is shown in Appendix \ref{appendix}. From Eq. \eqref{eq:u_R} we can directly deduce $\left|u_R\right|\le2$, which precisely gives rise to the positions of critical points. For the QWZ model, although the critical point at $u=0$ cannot be identified simply from the presence or absence of fidelity zeros, it is signaled by zeros approaching the real axis, mirroring the Lee-Yang paradigm of phase transitions.

\section{Conclusion}
In summary, we have investigated fidelity zeros in two-band topological models by extending the phase transition driving parameter into the complex plane. We unveiled that fidelity zeros emerge precisely when the real part of the energy gap closes for certain momentum modes, thereby establishing a direct connection between fidelity zeros and the spectral properties of the corresponding non-Hermitian Hamiltonian.
Using the Kitaev chain, the Haldane model, and the QWZ model as representative examples, we demonstrated that fidelity zeros exhibit universal distribution patterns. In finite-size systems, they appear as discrete lines parallel to the imaginary axis, with their real parts confined to a finite interval. Remarkably, the boundaries of the interval converge to the critical points of the underlying topological phase transitions in the thermodynamic limit.
Our study extends the fidelity-zero framework to topological quantum phase transitions, demonstrating its effectiveness as a diagnostic tool beyond symmetry-breaking transitions.

\section*{Acknowledgment}
This work is supported by National Key Research and Development Program of China (Grant No. 2023YFA1406704) and the NSFC under Grants No. 12474287, No. 12547107, and No. T2121001. 

\section*{Data Availability}
The data that support the findings of this article are not publicly available. The data are available from the authors upon reasonable request.

\appendix

\section{Verification of analytic expressions for fidelity zeros}
\label{appendix}

\begin{figure}
    \centering
    \includegraphics[width=0.8\linewidth]{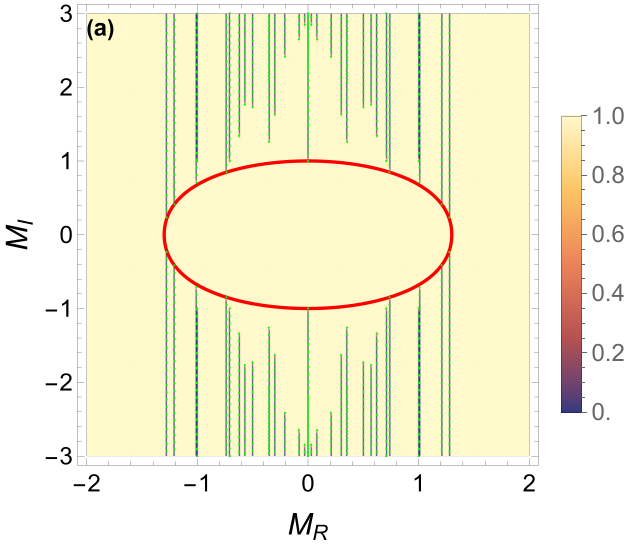}
    \includegraphics[width=0.8\linewidth]{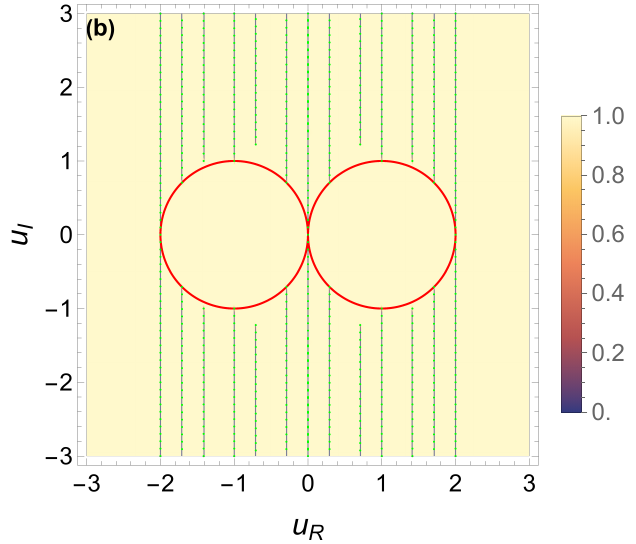}
    \caption{Comparison between fidelity zeros computed from the analytic expressions (green dots) and those obtained numerically (dark blue lines), for (a) the Haldane model and (b) the QWZ model with $L=8$. }
    \label{fig:appendix}
\end{figure}

In this section, we verify the analytic expressions of fidelity zeros for the Haldane model [Eq. \eqref{eq:Haldane FZ}] and the QWZ model [Eq. \eqref{eq:QWZ FZ}]. In Fig. \ref{fig:appendix}, The green dots are obtained using the analytic expressions, showing full agreement with the numerically obtained ones (dark blue lines).

\bibliography{bibtex}

\end{document}